# TiCkS: A FLEXIBLE WHITE-RABBIT BASED TIME-STAMPING BOARD

C. Champion[1], M. Punch[‡], R. Oger, S. Colonges, for the CTA Consortium,
APC, Univ Paris Diderot, CNRS/IN2P3, CEA/Irfu, Obs de Paris, Sorbonne Paris Cité, France,
and Y. Moudden, CEA – Cadarache / DRF / IRFM / STEP / GEAC

*Abstract*

We have developed the TiCkS board (Time and Clock Stamping) based on the White Rabbit (WR) SPEC node (Simple PCIe FMC carrier), to provide ns-precision time-stamps (TSs) of input signals (e.g., triggers from a connected device) and transmission of these TSs to a central collection point.

TiCkS was developed within the specifications of the Cherenkov Telescope Array (CTA) as one of the candidate TS nodes, with a small form-factor allowing its use in any CTA camera.

The essence of this development concerns the firmware in its Spartan-6 FPGA (Field-Programmable Gate Array), with the addition of: 1) a ns-precision TDC (Time-to-Digital Convertor) for the TSs; and 2) a UDP stack (User Datagram Protocol) to send TSs and auxiliary information over the WR fibre, and to receive configuration & slow control commands over the same fibre.

It also provides a PPS (Pulse Per Second) and other clock signals to the connected device, from which it can receive auxiliary event-type information over an SPI link (Serial Peripheral Interface).

A version of TiCkS with an FMC connector (FPGA Mezzanine Card) will be made available in the WR OpenHardware repository, so allowing the use of a mezzanine card with varied formats of input/output connectors, providing a cheap, flexible, and reliable solution for ns-precision time-stamping of trigger signals up to 400 kHz, for use in other experiments.

## CTA: CHERENKOV TELESCOPE ARRAY

The Cherenkov Telescope Array (CTA) [1] will be a gamma-ray observatory in the very-high-energy range (VHE, above around 30 GeV), consisting of over 100 imaging atmospheric Cherenkov telescopes (IACTs) distributed over two sites, one in each hemisphere; La Palma, Spain and the Atacama desert in Chile. The telescopes detect the few-nanosecond Cherenkov light flashes from the showers of particles (EAS, Extended Air Showers) initiated by gamma rays from cosmic sources, but also those produced by charged cosmic rays, which are thus background noise.

CTA will use a SoftWare Array Trigger (SWAT) to detect time coincidences – within a window up to 100 ns – between the signals from each telescope's Camera Trigger Management electronics (CTM). This allows the rejection of non-coincident images from Night-Sky Background light or isolated muons, and the "event-building" for stereoscopic events. Such a SWAT can be more flexible than a hardware trigger using transmission of analogue trigger signals and delay lines to correct for differential delays due to different sky-pointings.

The SWAT needs accurate relative TSs from each telescope's CTM, which it can correct in software for the telescope pointing directions, and identify coincidences within a flexible coincidence window, with a flexible topology and coincidence logic.

The White Rabbit (WR) technology has been adopted by CTA for this time-stamping. The TS relative accuracy for trigger coincidence identification for event-building is only at the tens of ns level. But, the timing information in the "wave-front" of Cherenkov photons hitting the array may contain further information, though likely redundant with the imaging information. Nonetheless, since the WR technology permits this, a 2 ns rms relative accuracy requirement was adopted CTA's timing nodes TSs.

## WHITE RABBIT TECHNOLOGY

White Rabbit [2] is an Open Hardware and Software project to provide sub-nanosecond synchronization accuracy combined with the flexibility and modularity of real-time Ethernet networks, based on timing synchronization over mono-mode fibres. It was initiated by CERN, the GSI Helmholtz Centre for Heavy Ion Research, and other partners from universities and industry starting in 2009. It is hoped that White-Rabbit will become a high-performance standard implementation of a future revised Precision Time Protocol.

For the purposes of CTA, WR permits to distribute the time from a central clock system to WR "nodes" in each telescope camera, over a hierarchical network of WR-compatible switches located at the array control centre. The WR-nodes time-stamp trigger signals from CTMs; both for "read-out" events for which there should be corresponding image data, and for "busy" triggers for those cameras which have dead-time (since the overall trigger pattern is useful at the event-reconstruction level).

Event and PPS counters in the camera's trigger electronics and its WR-node allow the image data to be combined with their time-stamps.

The White-Rabbit network itself may be used to collect the time-stamps from all telescopes at a central point, where the trigger coincidence logic can be implemented in a SoftWare Array Trigger (SWAT), and the coincidence information forwarded to each camera's data-processing pipeline to allow non-coincident trigger image data to be dropped.

## TiCkS: TIME & CLOCK STAMPING

We have developed the TiCkS board (Time and Clock Stamping) based on the WR SPEC node (Simple PCIe FMC carrier) [3].

The signals which are exchanged with the trigger electronics of the camera, using LVDS pairs (Low Voltage Differential Signalling) for transmission, follow the agreed CTA interface definition:

---

[1] email address     cedric.champion@apc.univ-paris7.fr
[‡] also at     Linnaeus University, Växjö, Sweden

- Camera → TiCkS:
  - Read-out Trigger signals
  - Busy Trigger Signals (*Optional*, only used for cameras with non-negligible dead-time)
  - (*Optional*) SPI (Serial Peripheral Interface) clock, data, and chip-select lines
- TiCkS → Camera
  - PPS signal (Pulse Per Second), synchronized with the central clock via White Rabbit
  - 10MHz clock, aligned with the PPS (for the cameras under test for this development, other cameras may use a different frequency)
  - External trigger signal, where the TiCkS trigger the camera at a defined TAI time.

This interface suffices to achieve the time-stamping and timing system monitoring functionality required by CTA.

*Hardware*

The SPEC node has been modified to reduce the form factor to 16.4×6.0×2.7 cm, removing the unused elements which are the PCI controller and connector, DDR3 memory, one of the SATA connectors, and – for the CTA-specific version – also the FMC connector which is replaced by two RJ45 jacks.

The power circuitry was adapted to allow 24V supply for the CTA-specific version. This only necessitated a change in the value of certain capacitors, inductances, and MOSFET, with the voltage regulator remaining the same. Thus, either the 12V or 24V supply version can be chosen when the board is stuffed. A more sturdy power connector (Buchanan terminal block) is chosen for the 30-year planned lifetime of CTA, for operation in cameras which reposition quickly.

The current drawn in operation is 160mA at 24V, so the power consumption is under 4W. The cost of a small serial production (~5) is under 500€ per board, including the SFP module interface to the fibre-optic cable.

*Firmware*

The firmware of the TiCkS is based on the standard WR core (v4.0 [4] was used in the tests below).

The WR-core provides a PPS and 125 MHz clock (referred to as "WR-clock" below) which are synchronized with the central clock over the fibre and WR switch link. So, natively, it can time-stamp signals with 8 ns precision.

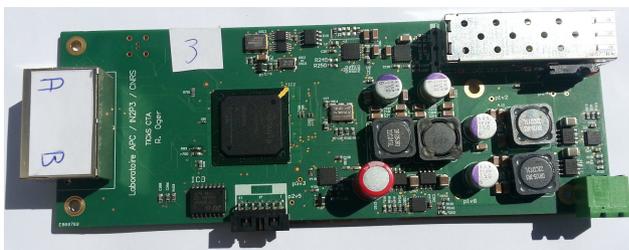

Figure 1: Photo of the TiCkS board. The two RJ45 connectors are on the left ("A" being the required timing connector, and "B" the optional busy and serial interface connector), with the cage for the SFP module on the right. The mini-USB connector is on the underside of the board.

To achieve ns-precision time-stamping of the input trigger signal, we use a 8-bit I/O-SerDes (Input/Output-SerializerDeserializer) clocked at 1 GHz SDR (Single Data Rate), where the 1 Ghz is generated by a Xilinx PLL from the WR 125 MHz clock. This is used as a shift register, and with some programmable logic resources for resynchronization purposes. We name this the "*fine-TDC*". The trigger signal from the CTM is fed into this fine-TDC, and in parallel is sent to the event counter (for which, the signal must be >20 ns to be counted).

A counter referred to as the "*coarse-TDC*" is used to keep track of the ticks of the WR-clock (every 8 ns). This counter is zeroed by the WR-PPS signal. At each tick of the WR-clock, the contents of the fine-TDC shift register are read, the presence of a rising front (reading MSB→LSB) in the register raises a flag that the fine-TDC has data, with the front position giving the number of nanoseconds since the last WR-clock tick. The flag triggers the read-out of the coarse-TDC counter together with the WR-provided International Atomic Time (TAI) seconds value, and a new flag that these data are ready is raised. These data constitute the full ns-precision TS, which is stored in a buffer. Finally, the coarse TDC is incremented whether or not there was a trigger.

On the subsequent tick of the Spartan-6 system clock (at 62.5 MHz), if there is a flag that the time-stamp is ready, the LSB (least-significant bit) parts of the data are put in a FIFO, where these are described in the "Data format, reception, and control" section below, and is resynchronized with the 62.5 MHz domain to avoid metastable states, after which the data-ready flag is lowered again. The FIFO used is one provided by the WR-core, with a width of 96-bit words (12 bytes) and a depth of 40 words (twice the depth of the packets of events which are sent, see below).

We found that a significant dead-time was introduced during the read-out of the 96-bit word FIFO to the 16-bit word UDP stack (6 clock cycles per event). To get around this, as a "fix" we use a MUX (multiplexer) to switch between two such FIFOs, in ping-pong mode.

The accompanying SPI data, currently 16-bits of event-type and related data, is awaited, and added to the word in the FIFO. The SPI-core used is that from OpenCores [5], running at 50 MHz, so it should take ~320 ns for the 16 bits of data to arrive from the CTM. Once it is ready, the SPI data register is added to the word in the FIFO, but in any case a time-out is implemented currently at 400 ns, after which the TiCkS allows new triggers to be input for time-stamping.

Note that this procedure is applied independently for both the read-out triggers and the busy triggers, with two instances of all TDCs and counters being implemented for this purpose.

Once the FIFO has reached 20 events, or after a predetermined time-out of 200 ms since the previous time it was emptied, then the full parts of the final TS (LSB and MSB, Most Significant Bits), and the full contents of the counters are added as a "*tailer*".

The resulting "bunch" of up to 20 events with their tailer is sent over the WR-fibre and WR-switches link to a predefined address using the UDP protocol. We use the

UDP stack from OpenCores [6]. The choice of UDP was made for simplicity and availability, with the consideration that UDP can reach higher throughput than TCP or more complex protocols, with faster connection setup since there is no initial handshake, and achieving faster packet delivery times, lower latency, and less risk of overloading the network by re-sending of lost packets (since UDP drops packets silently). But, there is the corresponding risk of some packet drop, which is mitigated by having the WR network on its own VLAN, and as shown below is negligible at the rates considered for CTA.

The firmware has the following summary characteristics, for the Spartan 6 `xc6slx45t-3fgg484` target device and version `ISE 14.7`, it is using 6735 slice registers (12% of the available resources), and one of the eight high-speed I/O buffers (`BUFPLL`), but all four of the available Phase Locked Loops (`PLL_ADVs`).

*Data Format, Reception, and Control*

**Data Format:** The data format was refined over several iterations in order to minimize the average event size. Individual event records contain only the Least Significant Bits (LSBs) of the different counter and TS fields, i.e., those which are susceptible to change within a bunch given its number of events and limit in time due to the time-out for sending bunches, allowing this event size to be reduced to 12 bytes. The 20-byte bunch tailer contains the full information for the counters and fields for the final event, with which each event can be unambiguously reconstituted.

If every event contained the full information, then 157 bits would be required (i.e., just under 20 bytes) plus 42 bytes of overhead. Whereas, packing the events in bunches of 20 gives a reduction of average event size to 15 bytes with overhead, so a reduction of 75%. This may be important since the WR links are currently at 1 Gbps.

A further advantage of grouping the events in bunches, is that it is strongly discouraged to have too-small packets (requiring CPU processing power for reception) or too-large (with higher probability of loss and fragmentation). A size of around 500 bytes is usually recommended, which is close to the 302-byte packets resulting from the 20-event bunches plus tailer, plus other overheads.

For a telescope at a maximum rate of 15 kHz, this gives a TS and associated data stream of 1.8 Mbps, which must traverse the 1 Gbps WR link, but at the uplink of the WR switch, it must share this with the other – up to 17 – telescopes on that switch. So the maximum rate could be 31 Mbps at that point, or only ~3% occupation (with an equivalent packet rate below 13000 packets/second).

Note that in a previous version of the WR-switch firmware (v3.3) the switch started to lose frames at high link loads, especially at low (~64-byte) payloads (see [7]). Though this problem has been resolved in the latest firmware versions, we consider it nonetheless prudent to have payloads of reasonable size combined with rates on the link which are far below 100%.

**Data Reception:** For the reception of the bunches, we have implemented a library in "c" which decodes the data bunches which arrive via UDP and reconstitute the events individually in a standard c structure, with no packing. This library can be used by whichever instance is required to receive the TSs and associated data.

For the tests described below, a process receives the TS bunch stream and either writes the TS data for later analysis, or produces a histogram of the distribution of times between successive events and successive event numbers. This has been essential for the initial debugging phase. The test-bench acquisition ran on a rather old PC with an Intel Core2 Duo Processor *E8600*.

Testing our decoding library gives that 13 ns are required for decoding, 34 ns if additionally each event is added to a histogram, and 583 ns if they are also all saved to disk. So, even old hardware should be able to treat up to 1.7 MHz of TS data, which is a factor >4 above the full CTA Southern array TS rate from all telescopes.

For integration tests with CTA cameras, we have implemented a process which receives the UDP TS bunch stream, decodes it into the c-structure per event, and sends it on over TCP to the Camera event builder. Note that the downstream links no longer have the WR 1 Gbps limit, but are rather at 10 Gbps, so there is no need to pack the events for best performance. This set-up has successfully operated for the integration tests with partial cameras of the MST-NectarCam and LST cameras (see [8] for the common CTM, named the Trigger Interface Board).

**Control:** The TiCkS board is controlled by sending it commands using the UDP stack.

For the initial configuration, it obtains its IP address from a DHCP server which should be available and configured on the network, otherwise it defaults to a fixed address. Then, it should be sent the MAC address to which it should send its bunches, while the corresponding IP address is computed from the TiCkS's own IP address by replacing the final 10 bits with a fixed value, and the port to which it sends these is a pre-defined fixed value. With this configuration, the TiCkS can send its bunches to a designated machine on the network.

Another command permits to send the TiCkS a precise time in TAI at which it will then emit an "external trigger" to the CTM. For simplicity of implementation, always aligned with the 8 ns precision of the WR clock.

Two further commands allow to ensure that the counters in both the CTM and the TiCkS are reset simultaneously following a well-defined state machine. The "reset" command to the TiCkS sets the event and PPS counters to zero, and maintains them in that state. The counters in the CTM can then be reset via the camera control, and the CTM placed in a "get-ready" state in which it awaits a (false) external trigger from the TiCkS before re-starting triggering. The "get-ready" command sent to the TiCkS then will cause it to send an external trigger just after the succeeding PPS, at which point both the CTM counters and the TiCkS counters will be synchronized at zero, and both will go into the "running" state. However, it should be relatively simple in software to determine counter offsets at a given time, e.g., after the PPS, so this procedure may not be required.

In a future version we plan to also add further commands, such as: to configure the IP address and port that the TiCkS sends the data; to set the wait time for SPI between events (or minimum time between triggers); and to set the time-out after which a non-full bunch is sent. The current defaults are reasonable for normal operation, but these changes should be quite easily implemented, as not affecting the delicate timing logic.

We plan to also implement a "throttle" to avoid the TiCkS overwhelming the WR system if it receives spurious triggers at a high rate, by requiring a minimum time since the start time of a bunch before allowing triggers for a new bunch, equivalent to setting maximum sustained rate.

Finally, we have noted that the newest version of the WR-core (from 4.0 upwards) now includes SNMP monitoring (Simple Network Management Protocol) as a standard, so that the node performance can be easily monitored using standard tools, including the WR node's PLL lock state, Return-Travel Time, TAI value, and many other useful parameters.

## MEASUREMENT AND TEST

### Test-bench Set-up

For testing the time-stamping and throughput capabilities of the TiCkS firmware, as well as the operation of the TiCkS board implementation, we use the test bench shown in Fig. 2.

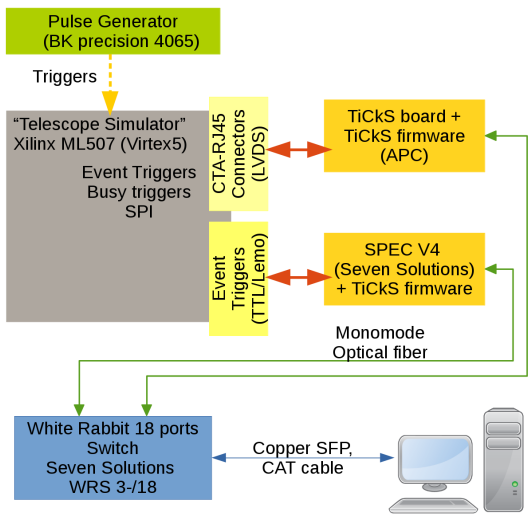

Figure 2: Set-up used for testing of the TiCkS board.

Two options were developed on this test-bench for the generation of event-trigger pulses:
1. Triggering from an external pulse generator with adjustable frequency, where the "Telescope Simulator" simply transforms the input to LVDS, adds SPI information, and fans-out these LVDS outputs simultaneously to a CTA-RJ45 mezzanine card and thence to the TiCkS board, and in simple TTL (Transistor-Transistor-Logic) for the event trigger information only, to an output connected to the SPEC board with its DIO (Digital I/O) mezzanine board.
- On the "Telescope Simulator", creating a random trigger with Poissonian distribution in time, with programmable average frequency (via the ML507's Dual-Inline Package – DIP – switches). This was implemented using the 20 LSBs of a 32-bit linear-feedback shift register (LFSR) clocked at 100 MHz, so giving a pseudo-random integer each 210 ns. This defines the minimum time and minimum interval. In order to simulate the operation of those cameras with dead-time, the trigger was sent on the "busy" output if the time since the last trigger was <380 ns, otherwise it was output on the "read-out" line.

- *Alignment of the PPS and 10MHz Outputs*

We have verified that the alignment between the PPS pulse and the 10 MHz clock output from the TiCkS board is small and stable over many cold restarts of the TiCkS board. This was achieved by creating a 200 MHz clock from the WR 125 MHz clock using a Xilinx PLL, and then dividing this clock down to 10 MHz with a simple counter, with this counter being started at the first PPS. This was taken care of since some cameras may rely on this for their internal timing. An example of this alignment is shown in Fig. 3, and is measured to be always below 1 ns.

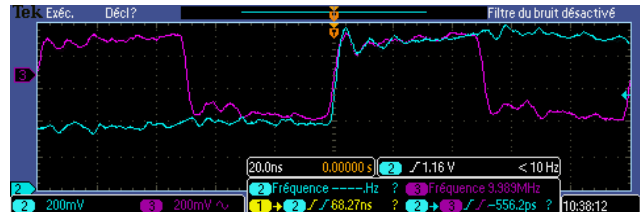

Figure 3: Check of the alignment between the PPS (blue trace) and the 10 MHz clock (purple trace) output from the TiCkS board.

### Maximum Event Rate Tests, and Missing Events

With the pulse generator input, and both TiCkS and SPEC being triggered, we have verified the response to fixed-frequency trigger inputs, by examining the histograms of times of successive events ($\Delta T_{intra}$). If a trigger is missed in this mode, it can be clearly seen since the resulting $\Delta T_{intra}$ will be twice the average value.

For increasing fixed trigger frequencies input, we measure on samples of a few $10^8$ triggers that there is no loss of events up to 320 kHz, with a slight loss at 400 kHz, and rising losses beyond that rate.

The corresponding distributions of $\Delta T_{intra}$ are not shown, since they mainly show the stability – within ±10 ns – of the pulse generator.

For the random triggers, some such $\Delta T_{intra}$ distributions are shown in Fig. 4, and are very well fitted by an exponential, as expected, with no obvious features. Looking for events which are seen in one card but not the other gives an incorrect estimate of the missed events since they will be correlated between the two cards.

A better estimate simply identifies skipped events, since the event counter is incremented even if the time-stamp cannot be formed because of dead-time or buffering loss in the time-stamping mechanism. This fractional loss is 0 at and below 9.5 kHz, and ~10 per million at 19 kHz which can be translated into an equivalent dead-time below 1 ns, which is negligible especially as a dead-time (~30 ns) is in any case required for a telescope to not re-trigger on the same shower.

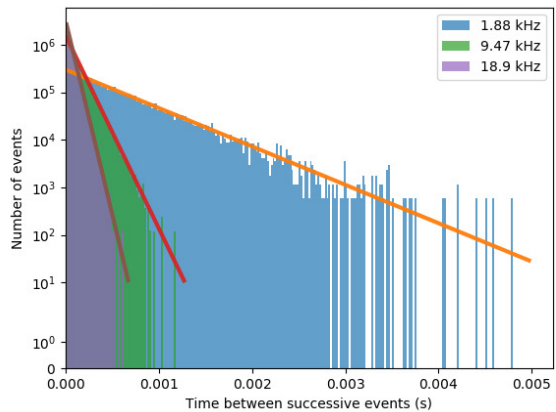

Figure 4: Time difference between successive injected events (random with Poisson distribution) for three typical average frequencies.

*Time Stamping Stability Between Boards*

For both of the test set-up options described above, we examined the distribution of TSs compared between the TiCkS and the SPEC ($\Delta T_{inter}$), using a simple coincidence window. This is shown in Fig. 5 for a fixed 10 kHz input frequency from the generator, both with and without an additional 62.3 ns cable delay. The distribution for a given measurement is confined to adjacent ns bins, and adding the cable delay shifts $\Delta T_{inter}$ as expected. For the randomly generated triggers, the distribution is similarly well-constrained.

Therefore, we conclude that the TiCkS hardware and firmware is well-capable of time-stamping input trigger signals with fixed or random time distribution, with little or no loss at the rates at which CTA will operate.

## CONCLUSIONS

We have developed and tested a TiCkS White Rabbit node based on the SPEC, with a scaled-down form factor and fewer components, with a firmware with additions to the WR core which allow ns-precision time-stamping to better than ns WR accuracy for two input channels ("read out" and "busy" triggers), the management of event and PPS counters for matching these time-stamps with the corresponding event data, and the transmission of these time-stamps, counters, and auxiliary data in a well-packed format to a defined address over the WR links.

We have added other functionalities such as SPI input for accompanying event types with accompanying minimum time between events.

We have shown that the TiCkS is capable of handling the rates expected for CTA in both fixed-frequency and random trigger input tests with negligible or no loss.

We plan to make the TiCkS firmware available both in a version which can be used on the TiCkS board with the CTA-defined 2xRJ45 connectors with LVDS signals, and also in a version which can be used on a SPEC board with the DIO mezzanine card, taking simple TTL signals (but the latter not including the SPI and Busy functionality).

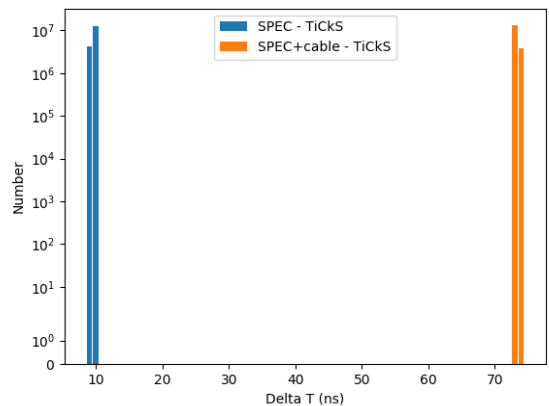

Figure 5: Time difference for injected events as measured by the TiCkS and SPEC WR-nodes. For the second measurement, a cable with measured transit time of 62.3 ns was added. For the random injected events, the distribution is similarly restricted to adjacent ns bins.

Together with this, we plan to make available the hardware version of the TiCkS board with an FMC connector rather than the 2xRJ45 CTA connectors.

These will soon be placed on the Open Hardware site where many other WR projects are hosted (see references) – once the reliability analysis has been completed for a hardened design reaching the goal of 15-year CTA operations between upgrades.

It will be available for the use of the CTA and wider communities, for whom such nanosecond time stamping in a low-cost, low-power package may be interesting.

## ACKNOWLEDGEMENT

Thanks to Julien Houles of the CPPM, Marseilles for support for the reception software of the TiCkS. This work was conducted in the context of the CTA Consortium Array Control Work Package. We gratefully acknowledge financial support from the agencies and organizations listed here: http://www.cta-observatory.org/consortium_acknowledgments .